\documentclass[twocolumn,notitlepage,nofootinbib,superscriptaddress,noeprint,aps,pra]{revtex4-2}
\pdfoutput=1
\usepackage[utf8]{inputenc}
\usepackage[english]{babel}
\usepackage{microtype}
\usepackage{amsmath,amssymb,bbm}
\usepackage[dvipsnames]{xcolor}
\usepackage[breaklinks=true,colorlinks=true,linkcolor=teal,urlcolor=teal,citecolor=teal]{hyperref}
\usepackage{orcidlink}

\newcommand{\id}{\mathbbm{1}} %identity operator

\newcommand{\stkout}[1]{\ifmmode\text{\sout{\ensuremath{#1}}}\else\sout{#1}\fi}

\newcommand{\ket}[1]{|#1\rangle} %ket
\newcommand{\bra}[1]{\langle#1|} %bra

\DeclareMathOperator{\Tr}{Tr}

\let\Re\relax
\let\Im\relax
\DeclareMathOperator{\Re}{Re}
\DeclareMathOperator{\Im}{Im}

\renewcommand{\vec}{\boldsymbol}

\begin{document}

\title{Quantum illumination advantage in quantum Doppler radar}

\author{Rongyu Wei\,\orcidlink{0000-0001-9721-8732}}
\email{rongyu.wei@qq.com}
\affiliation{National Key Laboratory of Radar Signal Processing, Xidian University, Xi'an, 710071, China}
\affiliation{Scuola Normale Superiore, I-56126 Pisa, Italy}

\author{Francesco Albarelli\,\orcidlink{0000-0001-5775-168X}}
\email{francesco.albarelli@gmail.com}
\affiliation{Scuola Normale Superiore, I-56126 Pisa, Italy}

\author{Jun Li}
\email{junli01@mail.xidian.edu.cn}
\affiliation{National Key Laboratory of Radar Signal Processing, Xidian University, Xi'an, 710071, China}

\author{Vittorio Giovannetti\,\orcidlink{0000-0002-7636-9002}}
\email{vittorio.giovannetti@sns.it}
\affiliation{Scuola Normale Superiore, I-56126 Pisa, Italy}

\begin{abstract} 
A Doppler radar is a device that employs the Doppler effect to estimate the radial velocity of a moving target at a distance.
Traditional radars are based on a classical description of the electromagnetic radiation, but in principle their performance can be improved employing entangled quantum probe states.
For target detection, i.e. hypothesis testing, a quantum advantage exists even in the high-noise regime appropriate to describe microwave fields, a protocol known as quantum illumination.
In this paper, we show a similar advantage also for a quantum Doppler radar operating in presence of thermal noise, whereas so far a quantum advantage was shown in the noiseless scenario or in lidars operating at optical frequencies with negligible thermal noise.
Concretely, we quantify the radar performance in terms of the quantum Fisher information, which captures the ultimate precision allowed by quantum mechanics in the asymptotic regime.
We compare a classical protocol based on coherent states with a quantum one that uses multimode states obtained from spontaneous parametric downconversion.
To ensure a fair comparison we match the signal energy and pulse duration.
We show that a 3dB advantage is possible in the regime of small number of signal photons and high thermal noise, even for low transmissivity.
\end{abstract}

\maketitle

%%%%%%%%%%%%%%%%%%%%%%%%%%%%%%%%%
\section{Introduction}

Quantum technologies seek to take advantage of genuine quantum effects for increased performances, ideally beating any classical counterpart.
While the field is growing and steadily advancing towards concrete applications, adapting theoretical ideas to nonidealities of the real world remains challenging.
From a physical perspective, quantum optics is one of the cornerstones of modern quantum technologies; indeed, many tasks in which the electromagnetic field is used as a probe, or as the carrier of information, show the potential for a quantum advantage.
In this regard, quantum metrology~\cite{Giovannetti2011,Polino2020} and quantum sensing~\cite{Pirandola2018} are among the most promising applications, including quantum-enhanced protocols for, e.g., interferometry~\cite{Demkowicz-Dobrzanski2015a}, spectroscopy~\cite{Dorfman2016}, superresolution~\cite{Tsang2019a} and imaging~\cite{Genovese2016,Defienne2024}.

One of the sensing tasks that has attracted more interest from physicists and engineers in the past decade is remote detection of targets, leading to ideas for quantum lidar and radar applications.
% Indeed, useful real-world applications already exist, especially in the context of optical interferometry, e.g. exploiting quantum squeezing for increasing the sensitivity of gravitational wave detection.
% The theoretical research for quantum
% technologies such as lidar 
% and it has already shown useful applications in the optical domain.
In this context, the most paradigmatic and studied protocol, known as quantum illumination (QI)~\cite{Lloyd2008,Tan2008,Shapiro2009}, concerns the detection of a weakly reflecting target in a noisy thermal environment: a signal beam entangled with an idler leads to increased success probability with respect to any classical beam.
After the first proposal, a lot of efforts have been devoted to concretize this protocol.
There have been experimental demonstrations of quantum advantage in QI at optical frequencies~\cite{Zhang2015e,Xu2021d,Hao2022a}.
However, the real challenge is to apply the QI protocol in the microwave band, since in this regime the thermal noise is naturally very high.
Theoretical efforts started with Ref.~\cite{Barzanjeh2015}, and new promising design for microwave receivers have been proposed recently~\cite{Shi2024e,Angeletti2023,Reichert2023}.
Several experiments in the microwave domain have been performed~\cite{Luong2018,Chang2019,Barzanjeh2020}, but only recently a real advantage was obtained at cryogenic temperature~\cite{Assouly2023}.
Progress in these fields has been recently surveyed~\cite{Shapiro2020,Torrome2021,Sorelli2022,Casariego2023,Karsa2024,Torrome2024,Zhang2024x}.

Going beyond detecting the presence of a target, classical lidar and radar systems are also used to estimate properties of the target, most notably its distance, i.e. ranging, and its radial velocity via Doppler effect.
Assuming negligible losses and thermal noise, range and velocity measurements can be regarded as a unitary estimation problem and thus it is possible to reach the Heisenberg limit, a mean squared error inversely proportional to the number of signal photons, representing a quadratic improvement with respect to the performance of classical strategies.
The estimation of range~\cite{Giovannetti2001a,Giovannetti2002,Shapiro2007,Maccone2020b}, velocity~\cite{Reichert2022} or both parameters simultaneously~\cite{Zhuang2017b,Huang2021,Reichert2024} has been studied in this noiseless regime, which may only be appropriate for optical frequencies and short distances.

Reaching the Heisenberg limit represents the most dramatic advantage in quantum metrology and sensing, but it disappears asymptotically due to most kinds of noise~\cite{Demkowicz-Dobrzanski2012}, even using the most sophisticated strategies to counteract its effect~\cite{Kurdzialek2023a}.
This applies to most real scenarios, including ranging and velocity estimation in presence of optical losses.
Nonetheless, even if an asymptotic quadratic advantage is ruled out by the noise, the precision in estimating parameters can still be enhanced by a constant factor.
In the microwave domain, not only losses, but thermal noise is naturally present in real applications.
The advantages of quantum strategies for range estimation in this scenario have been analyzed in Refs.~\cite{Zhuang2021,Karsa2021a,Zhuang2022,Cohen2022a}.
However, to the best of our knowledge, no studies have addressed a quantum Doppler radar in the microwave domain, considering the effect of thermal noise.

In this paper we fill this gap by showing that a 3dB advantage is possible for a Doppler radar, in the same high-thermal noise, high-loss and low-signal-power regime where the original QI protocol is effective.
More in detail, in our results we consider the quantum Fisher information (QFI) as a figure of merit for the comparison between a classical strategy employing a coherent state and a quantum strategy using an entangled state, obtained by pulsed spontaneous parametric downconversion (SPDC).
For a given quantum state, the QFI captures the maximal amount of information about the parameter which can be extracted if an optimal measurement is implemented, in the limit of many repetitions of the experiment.

Crucially, in order to fully describe the Doppler effect, including how it affects the spectrum, one has to deal explicitly with a continuous-frequency description of the involved fields~\cite{Huang2021,Reichert2022}.
Therefore, the problem requires a more involved model than the original QI scenario for target detection, where a simple description of the signal evolution in terms of a single-mode bosonic channel suffices~\cite{Tan2008,Shapiro2020}.
Following the approach of Ref.~\cite{Reichert2022} we perform calculations by treating the continuum field in terms of discrete modes.
In this setting, we deal with the estimation of a parameter that affects the underlying structure of the modes of the electromagnetic field, a task for which specific techniques have recently been developed~\cite{Gessner2023a,Sorelli2024}.

The paper is structured as follows.
In Sec.~\ref{sec:model} we introduce the model, including the discrete-mode description and the two classes of classical and quantum probe states.
In Sec.~\ref{sec:results} we explain our approach for comparing the performance of classical and quantum strategies and we show numerical results.
Sec.~\ref{sec:conclusions} concludes the main part of the paper with some remarks and discussion of open problems.
Sec.~\ref{sec:methods} contains most of the technical details, including an introduction to the theoretical framework for quantum estimation theory and Gaussian quantum systems, as well as more details on the calculation of the QFI in the two cases.
\begin{figure}
  \centering
  \includegraphics[width=.75\columnwidth]{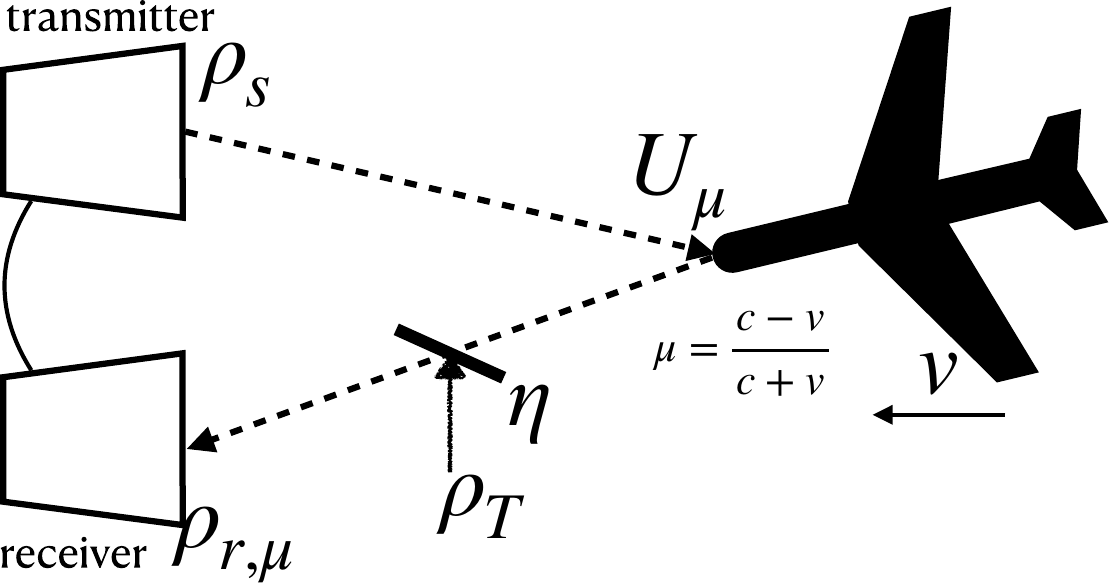}
  \caption{Schematic representation of the quantum Doppler radar with thermal noise.
  The Doppler parameter $\mu$ is encoded during the reflection at the target, while the action of thermal noise is modeled as a beam splitter interaction with frequency-independent transmissivity $\eta$.
  }
  \label{fig:scheme}
\end{figure}

%%%%%%%%%%%%%%%%%%%%%%%%%%%%%%%%%
\section{Model}
\label{sec:model}

We consider a signal beam in a single spatial and polarization mode within the paraxial approximation, so that the electromagnetic field can be considered one-dimensional.
Thus, the field can be described in terms of a single continuous frequency~\cite{Blow1990} through the bosonic operators $\hat{a}_{S}(\omega)$ (units of $1/\sqrt{\mathrm{frequency}}$), satisfying the commutation relations $[ \hat{a}_{S}(\omega) , \hat{a}^\dag_{S}(\omega') ] = \delta(\omega- \omega')$.
One of the main parameters of interest for QI and radar applications is the average number of photons in the signal beam, defined as
\begin{equation}
  \label{eq:NSgeneral}
  N_S = \int_0^\infty d \omega \Tr \left[ \rho_s \hat{a}^\dag_S(\omega) \hat{a}_S(\omega) \right],
\end{equation}
where $ \rho_s$ is the initial (transmitted) state of the signal beam.
For the quantum protocol, similarly to QI, we also include an idler beam in an orthogonal spatial mode that does not interact with the target and remains stored at the receiver, corresponding to a second one-dimensional bosonic field, with operators $\hat{a}_{I}(\omega)$.

\subsection{Doppler parameter encoding and noise model}

We model the target as a perfectly reflective mirror, in motion towards the radar with constant radial velocity $v$; absorption of radiation by the target can be incorporated later together with the propagation losses.
We adopt the description of the Doppler effect introduced in Refs.~\cite{Huang2021,Reichert2022}, which is more accurate than modeling it just as a frequency shift as in Ref.~\cite{Zhuang2017b}, since it includes effects on the frequency spectrum.
The reflection of the signal beam by the moving target acts in the Heisenberg picture as a linear transformation of modes
\begin{equation}
\label{eq:DopplerUmu}
  \hat U_\mu ^\dag \hat a_{S}(\omega ){\hat U_\mu } =  - {\mu ^{1/2}}\hat a_{S}(\mu \omega)
\end{equation}
where we adopt the convention that the target is moving towards the radar with radial speed $v$; we have introduced the quantity $\mu  = \frac{{c - v}}{{c + v}}$, which will be our parameter of interest in the estimation task.
This transformation is passive, since it preserves the number of photons $N_S$, while the mean energy $ \hbar \int \! d \omega \, \omega  \langle \hat{a}^\dag_S(\omega) \hat{a}_S(\omega) \rangle$ is not conserved, since the frequency spectrum is modified.

As a simple example, and to check that this map acts as expected, we consider a single-photon state $ \ket{1_f} = \int d\omega f(\omega) \hat{a}_S ( \omega)^\dag \ket{0}$ with spectral amplitude $f(\omega)$.
Eq.~\eqref{eq:DopplerUmu} then implies that the spectral density $|f(\omega)|^2 = \bra{1_f} \hat{a}_S(\omega)^\dag \hat{a}_S(\omega) \ket{1_f}$, obtained in this case by frequency-resolved photon counting, transforms as $ |f(\omega)|^2 \rightarrow - \mu^{-1} |f(\omega \mu)|^2$: the mean $\omega_c = \int \! d \omega \, \omega |f(\omega)|^2 $ is shifted to $\omega_c / \mu$ and the bandwidth $\Delta \omega = \left[ \int \! d\omega \, (\omega - \omega_c)^2 |f(\omega)|^2  \right]^{1/2}$ is mapped to $ \Delta \omega / \mu$.
For $ v > 0$ (target approaching the radar) $\mu < 1$, the field mean frequency is blueshifted as expected while the spectral distribution is broadened.

A complete description of the received field, in the noiseless case, includes also the phase factor $\exp \left[ {i\left( {\omega/\mu - \omega } \right){t_0} + i (\omega/\mu) \tau } \right]$, where $\tau$ is the roundtrip time.
If the transmitter and the receiver are at the same position, then $\tau  = \frac{{2x}}{{c + v}}$ with $x$ representing the distance from the transmitter to the target (range).
This unitary transformation gives the complete dependence on range and velocity, as considered in Ref.~\cite{Huang2021} for the single-photon sector.
Even if the phase contains information on the velocity $v$, we will neglect this contribution and only focus on the effects pertaining to the frequency spectrum.
The reason for doing so is that acquiring phase information requires additional knowledge of the target's distance, and that in practical scenarios the phase is often randomized~\cite{Reichert2022}.

We model the action of the thermal noise as a beam splitter interaction, assuming that the transmissivity $\eta$ does not depend on the frequency.
This induces the total transformation on the signal modes
\begin{equation}
  \label{eq:full_noisy}
      \begin{aligned}
     {{\hat a}_R}(\omega )  & = \hat U_{\eta ,b}^\dag \hat U_\mu ^\dag {{\hat a}_S}(\omega ){{\hat U}_\mu }{{\hat U}_{\eta ,b}}\\
   &=   - {\mu ^{1/2}}\sqrt \eta  {{\hat a}_S}(\mu \omega ) + \sqrt {1 - \eta } \hat b(\omega )
   \end{aligned}
\end{equation}
here the field $\hat b(\omega )$ is initially in a thermal state $\rho_T$ with average photon number ${{ \bar{n}_B}(\omega )}/{(1 - \eta )}$, which satisfies $\Tr[ \rho_T \hat{b}^\dag(\omega) ] = 0$ and $\Tr[ \rho_T \hat{b}^\dag(\omega') \hat{b}(\omega) ] = \delta(\omega-\omega') \bar{n}_B(\omega) /{(1 - \eta )}$.
This loss-dependent rescaling of the mean thermal occupation number is a typical assumption in QI, which removes the possibility of detecting a target from its shadow: if no signal beam is sent towards the target only thermal noise is received.
In the regime $\eta \ll 1$ of interest for QI, this correction is expected to be innocuous since $(1 - \eta ) \approx 1$; a critical discussion of this assumption is offered in Ref.~\cite{Volkoff2024}.

An important observation concerns the frequency dependence of the mean number of thermal photons ${ \bar{n}_B}(\omega )$.
In principle, modelling the noisy thermal environment as an ideal blackbody, the density of photons should be consistent with Planck's law for the energy density.
Nonetheless, we also make a quasimonochromatic (narrowband) approximation for the probe radiation, meaning that it has a spectrum peaked around a central frequency $\omega_c$ and $\Delta \omega \ll \omega_c$.
This implies that we can safely consider the number of thermal excitation as a constant $N_B \equiv \bar{n}_B(\omega_c) =  \left( \exp[ \hbar \omega_c / ( k_B T) ] -1 \right)^{-1}$; this is a standard assumption, see e.g. Refs.~\cite{Zhuang2022,Reichert2023}.
The narrowband assumption also justifies the initial choice of a constant $\eta$.

In a more realistic description, the thermal noise is mixed with the signal beam both during the propagation from the transmitter to the target, and during the propagation back to the receiver.
The equivalence to the effective description in Eq.~\eqref{eq:full_noisy} in terms of a single thermal field, schematically represented in Fig.~\ref{fig:scheme}, is guaranteed because we make the natural assumption that the temperature is the same for the two propagations, together with the narrowband approximation.
In principle, the application of $\hat{U}_\mu$ before or after the interaction with the noise gives a different result in terms of mode transformations.
However, the thermal field is not measured at the end, and it has zero first moments, so it affects the final result only through its fluctuations, which we assume to be frequency-independent in the narrowband limit.
In other words, the action of the noise corresponds to a thermal channel whose action commutes with the Doppler unitary (this will also be clear from the discrete mode description in the next section).
Under these assumptions, the two channels describing the two legs of the trip, with transmissivities $\eta_1$ and $\eta_2$ respectively, can be combined into a single thermal channel with $\eta = \eta_1 \eta_2$.

\subsection{Discrete modes}

Instead of considering a continuous-frequency description of fields we can introduce a discrete basis of orthonormal modes~\cite{Blow1990,Rohde2007} $\{ \Psi_k \}_{k=0}^{\infty}$, i.e. orthonormal square integrable functions 
\begin{equation}
    \int d \omega \, \Psi_k^*(\omega) \Psi_l(\omega) = \delta_{lk}
\end{equation}
that also satisfy the completeness relation
\begin{equation}
    \sum_{k=1}^{\infty} \Psi_k^*(\omega) \Psi_k(\omega') = \delta(\omega - \omega').
\end{equation}
We can also introduce the corresponding discrete bosonic annihilation (creation) operators, which destroy (create) an excitation in the mode $\Psi_k$:
\begin{equation}
    \hat{A}_{S,k} = \int d \omega_S \, \Psi_k(\omega_S) \hat{a}_S(\omega),
\end{equation}
and the same can be done for the idler modes.
The discrete bosonic operators satisfy $[\hat{A}_{j} , \hat{A}_{k}^\dag ] = \delta_{jk}$ and will be denoted by capital letters.
The completeness property guarantees that we can also invert the relation between the continuous and discrete description as $\hat{a}(\omega) = \sum_k \Psi_k^*(\omega) \hat{A}_k$.
In the following, we will make different choices for the mode basis $\{ \Psi_k \}$ depending on the initial state, with the aim to simplify the calculations.

The Heisenberg-picture evolution of the discrete modes can be obtained by integrating Eq.~\eqref{eq:full_noisy} with the spectral amplitudes of the basis functions: 
\begin{equation}
  \label{eq:Doppler_mode_transf_discrete}
 \hat{A}_{k,r} \mapsto \sqrt{\eta} \, \hat{U}_\mu^\dag \hat{A}_{k,s} \hat{U}_\mu + \sqrt{1-\eta} \, \hat{B}_{k},
\end{equation}
where $\hat{B}_{k} = \int d \omega \, \Psi_k(\omega) \hat{b}(\omega)$ are modes of the noise field.
At the discrete level, the unitary parameter encoding corresponds a unitary parameter-dependent reshuffling of the modes
\begin{equation}
  \label{eq:Doppler_Heisenberg_discrete}
  \begin{split}
  \hat{U}_\mu^\dag \hat{A}_{k,s} \hat{U}_\mu &= \int \! d \omega_S \, \psi_k(\omega_S) \hat{U}_\mu^\dag \hat{a}_{S}(\omega_S) \hat{U}_\mu  \\
 & = -\mu ^{1/2} \int \! d \omega_S \, \psi_k(\omega_S)  {{\hat a}_S}(\mu \omega_S ) \\ 
 & = -\mu ^{1/2} \int \! d \omega \, \psi_k(\omega)   \sum_j \psi_j( \mu \omega ) \hat{A}_j \\ 
 & = \sum_j \left( {-\mu ^{1/2}} \int d \omega \psi_k(\omega)  \psi_j( \mu \omega ) \right) \hat{A}_j \\
 & = \sum_j \mathcal{U}_{\mu,kj} \hat{A}_j,
%  {{\hat a}_S}(\mu \omega )
  \end{split}
\end{equation}
where $\mathcal{U}_{\mu,kj}$ are elements of a unitary matrix.
Thanks to the quasimonochromatic approximation, we have that each discrete mode in the basis interacts with an independent but identical environment, i.e. 
\begin{equation}
  \Tr\left[ \rho_B \hat{B}_{k}^\dag \hat{B}_{k} \right] = 
  \int d \omega \frac{| \Psi_k( \omega ) |^2}{ e^{\frac{ \hbar \omega_c}{k T}} - 1 }  = \bar{n}_B(\omega_c) = N_B
\end{equation}

In Schrödinger picture\footnote{In Appendix~\ref{app:schrodinger_picture_propagating_fields} we present a pedagogical comment about using Schrödinger picture for the description of a propagating electromagnetic field.}, the initial transmitted state $\rho_s$ undergoes a bosonic Gaussian channel, defined by tracing away the noise:
\begin{equation}
  \rho_{r,\mu} = \Tr_b \left[ \hat U_{\eta ,b} \hat{U}_\mu \left( \rho_s  \otimes \rho_T  \right) {{\hat U}_\mu^\dag }{{\hat U}_{\eta ,b}^\dag}  \right],
\end{equation}
which gives the $\mu$-dependente received state $\rho_{r,\mu}$.
The transformation from the transmitted stare $\rho_s$ to the received state $\rho_{r,\mu}$ is schematically shown in Fig.~\ref{fig:scheme}. 
In absence of the Doppler effect  $\mu = 1$ (so that $\mathcal{U}_{\mu,kj} = \delta_{k,j}$) the channel is a multimode thermal channel with the same temperature for all the modes.
Our figure of merit will be the QFI of the received state $\rho_{r,\mu}$ with respect to the parameter $\mu$, for different choices of the initial signal state $\rho_s$, which may also include an additional idler beam.

\subsection{Classical light}

A coherent state in the continuous-frequency setting is defined as $\ket{ \alpha_f }  = \hat D_f(\alpha ) \ket{  0  } $ in terms of the displacement operator
\begin{equation}
\label{eq:displ1}
    \begin{aligned}
    \hat D_f(\alpha ) =&  \exp \left[ {\alpha \int {d\omega f(\omega )\left( {\hat a(\omega ) - {{\hat a}^\dag }(\omega )} \right)} } \right] \\ 
    = & \exp \left[ {\alpha \left( {{{\hat A}_f} - \hat A_f^\dag } \right)} \right].
    \end{aligned}
\end{equation}
Here $f(\omega)$ is a square-integrable normalized wave function that defines the single discrete mode populated by the coherent state (we assume $f$ to be real-valued), corresponding to the bosonic operator $ {\hat A_f} = \int {d\omega } f(\omega )\hat a(\omega )$ that
% The corresponding \emph{wavepacket} bosonic operator
% In particular, we can notice that this state occupies a single temporal mode:
% If we define that 
% \begin{equation}
% \label{eq:Acohrent}
    % {\hat A_f} = \int {d\omega } f(\omega )\hat a(\omega )
% \end{equation}
annihilates excitations in this mode.
% The displacement can alternatively be written as
% \begin{equation}
%     \hat D_f(\alpha ) = 
% \end{equation}
% and 
For a coherent state the average photon number $N_S$ in Eq.~\eqref{eq:NSgeneral} is
\begin{equation}
  \label{eq:NS_coh}
N_S = \langle \alpha_f | \hat{A}_f^\dag \hat{A}_f | \alpha_f \rangle = \alpha^2.
\end{equation}
Since we are in the narrowband approximation, all frequency integrals can be extended to $-\infty$ and we can safely introduce the time-domain amplitude $\tilde{f}(t) = \frac{1}{{\sqrt {2\pi } }}\int {f(y){e^{iyt}}d} y$.
Then, the time duration $\Delta T$ of the pulse is given by the standard deviation of the modulus squared of the time-domain amplitude:
\begin{equation}
  \label{eq:DeltaT_coh}
  \Delta T^2 = \int \! dt \, t^2 |g(t)|^2 - \left( \int \! dt \, t |g(t)|^2 \right)^2 .
\end{equation}
The physical meaning comes from the time-domain bosonic operators $\hat{a}_S(t) = (2 \pi)^{-1/2}\int_{-\infty}^\infty e^{-i \omega t } a_S(\omega)$, which can be introduced in the narrowband approximation~\cite{Blow1990}.
Then $\hat{a}_S(t)^\dag \hat{a}_S(t)$ is the photon flux operator, and we can define the normalized mean photon flux $\mathfrak{f}(t)=\bra{\Psi_{\mathrm{SPDC}} } \hat{a}^\dag_S(t) \hat{a}_S(t) \ket{\Psi_{\mathrm{SPDC}} }/N_S$. 
For a coherent state this is simply $\mathfrak{f}(t)=|g(t)|^2 $, which gives an operational meaning to the definition of $\Delta T$ above.

In absence of thermal noise, the evolved state remains coherent (and thus in a single discrete mode) under the Doppler transformation in Eq.~\eqref{eq:DopplerUmu}, since $\hat{U}_\mu D(\alpha) U_\mu^\dag =  \exp \left[ {\alpha \left( { {\hat A}_{f,\mu}  - \hat A_{f,\mu}^\dag } \right)} \right]$ where $A_{f,\mu} = U_\mu A_{f} U_\mu^\dag$.
Analogously to the single-photon case discussed below Eq.~\eqref{eq:DopplerUmu}, the only effect is to transform the spectral amplitude to $-\mu^{-1/2} f(\omega \mu)$.
Moreover, losses do not affect the purity of coherent states, but just decrease the amplitude by a factor $\sqrt{\eta}$. 
For this reason the QFI calculation in absence of thermal noise is straight-forward~\cite{Reichert2022}.

For $N_B >0$ the calculation of the QFI is conceptually slightly more involved.
While the state remains Gaussian and standard formulas for the QFI of Gaussian states are well known~\cite{Serafini2023}, a proper choice of the modes used for the calculation is crucial.
Indeed, since the parameter enters the definition of the bosonic modes, one should in principle perform the calculation in a different parameter-independent orthonormal basis.
Luckily, it is enough to consider modes defined in terms of the $\mu$-derivatives of the originally populated modes~\cite{Gessner2023a,Sorelli2024}, an explicit justification of this fact is provided in Sec.~\ref{subsec:QFIcalc}.
Since in this case the initial state is single-mode, only one additional mode is needed for the QFI calculation.

\subsection{Quantum light}

We consider a quantum state that can be produced experimentally by spontaneous parametric down conversion (SPDC)~\cite{Grice1997,Mauerer2009,Mosley2009,Christ2011}.
The effective Hamiltonian describing this process is
\begin{equation}
    \label{eq:H_SPDC_cont}
    {\hat H_{\mathrm{SPDC}}} = i\hbar \xi \int {d{\omega _S}\int {d{\omega _I}} } f({\omega _S},{\omega _I})\hat a_S^\dag ({\omega _S})\hat a_I^\dag ({\omega _I}) + \mathrm{h.c.} \,;
\end{equation}
$\xi$ is a coupling constant that depends on the pump intensity and on the strength $\chi^{(2)}$ of the nonlinear susceptibility, which also takes into account the length of the nonlinear crystal.
We will refer to $\xi$ as the squeezing parameter, and it is chosen to be real-valued for simplicity.
The joint spectral amplitude (JSA) of signal and idler modes $f({\omega _S},{\omega _I})$ is defined to be square normalized, i.e. $\int d \omega_S \omega_I  |f({\omega _S},{\omega _I})|^2 = 1$.  
The SPDC state is defined as 
\begin{equation}
  \ket{\Psi_{\mathrm{SPDC}}} = e^{- \frac{i}{\hbar} \hat{H}_{\mathrm{SPDC}} } \ket{0},
\end{equation}
which corresponds to the state when the pulse has completely left the crystal and the interaction is finished.

In order to proceed with concrete calculations, it is fundamental to work with discrete modes, which can be achieved via the Schmidt decomposition of the function $f(\omega_S,\omega_I)$, i.e.
\begin{equation}
    \label{eq:Schmidt_dec}
    f({\omega _S},{\omega _I}) = \sum\limits_{m = 0}^M {{r_m}{\varphi _m}({\omega _S}){\phi _m}({\omega _I})},
\end{equation}
which is defined in terms of orthonormal polynomials $ \int d \omega \varphi_m(\omega) \varphi_n(\omega)  = \delta_{n,m}$, which are also complete $ \sum_m \varphi_m(\omega) \varphi_m(\omega') = \delta(\omega - \omega') $; the same properties hold for the idler modes. 
The weights are normalized $\sum {r_m^2}  = 1$.

Concretely, we will consider a double Gaussian JSA
\begin{equation}
  \begin{split}
  \label{eq:JSA_Gaussian}
  f({\omega _S},{\omega _I}) = & \sqrt {\frac{2}{{\pi \sigma_p \varepsilon }}} \exp \left( { - \frac{{{{({\omega _S} + {\omega _I} - {\omega _p})}^2}}}{{2{\sigma_p ^2}}}} \right) \\ 
  & \times \exp \left( { - \frac{{{{({\omega _S} - {\omega _I})}^2}}}{{2{\varepsilon ^2}}}} \right),
  \end{split}
\end{equation}
for which the corresponding Schmidt modes are Hermite-Gauss functions.
Notice that the first function represents the pump envelope, where $\omega_p $ is the pump frequency, and the signal will have a central frequency $\omega_c = \omega_p/2$.
The second term represents the phase-matching function and it is often well approximated by a Gaussian~\cite{Christ2011}.
For this Gaussian JSA we have the Schmidt weights
${r_m} = \frac{{2\sqrt {\sigma_p \varepsilon } }}{{\sigma_p  + \varepsilon }}{\left( {\frac{{\sigma_p  - \varepsilon }}{{\sigma_p  + \varepsilon }}} \right)^m}$
and we can also introduce the Schmidt number  $K = \frac{{{\sigma_p ^2} + {\varepsilon ^2}}}{{2\sigma_p \varepsilon }}$, which roughly captures the number of active modes.

Unlike coherent states, SPDC light intrinsically requires a multimode description, unless the function $f(\omega_S,\omega_I)$ factorizes, i.e. $r_m = \delta_{0,m}$ in Eq.~\eqref{eq:Schmidt_dec}.
Introducing the wavepacket creation operators for the Schmidt modes, i.e. $\hat A_{S,m}^\dag  = \int {d{\omega _S}{\varphi _m}({\omega _S})\hat a_S^\dag ({\omega _S})}$ and $\hat A_{I,m}^\dag  = \int {d{\omega _I}{\phi _m}({\omega _I})\hat a_I^\dag ({\omega _I})}$ we can rewrite the Hamiltonian as
\begin{equation}
  \label{eq:H_SPDC_disc}
    {\hat H_{\mathrm{SPDC}}} \propto \sum\limits_{m = 0}^M {{r_m}\hat A_{S,m}^\dag \hat A_{I,m}^\dag }  + h.c.
\end{equation}
In this description, the SPDC state has a simple product structure
\begin{equation}
  \label{eq:Psi_SPDC_disc}
  \ket{\Psi_{\mathrm{SPDC}} } = \bigotimes_{n=0}^{\infty} \frac{1}{\cosh (\xi r_n )} \exp\left[ -\tanh (\xi r_n) \hat{A}_{S,n}^\dag \hat{A}_{I,n}^\dag  \right] \ket{0},
\end{equation}
meaning that different Schmidt modes are uncorrelated, while each signal and idler pair in the Schmidt decomposition is populated by an entangled two-mode squeezed vacuum state.
We can also write expressions for the average number of signal photons 
\begin{equation}
  \label{eq:NS_SPDC}
  N_S = \sum_{m=0}^\infty N_{S,m} = \sum_{m=0}^\infty \sinh^2(\xi r_m),
\end{equation}
where $ N_{S,m} = \bra{\Psi_{\mathrm{SPDC}} } \hat{A}_{S,m}^\dag\hat{A}_{S,m} \ket{\Psi_{\mathrm{SPDC}} }$ is the average photons in each signal Schmidt mode.
Exploiting the flux operator for the signal, defined below Eq.~\eqref{eq:DeltaT_coh}, the time duration of the signal is defined in the same way through the normalized mean photon flux $\mathfrak{f}(t)=\bra{\Psi_{\mathrm{SPDC}} } \hat{a}^\dag_S(t) \hat{a}_S(t) \ket{\Psi_{\mathrm{SPDC}} }/N_S$.
For the Gaussian joint spectral amplitude in Eq.~\eqref{eq:JSA_Gaussian} the time duration has the explicit expression
\begin{equation}
  \label{eq:DeltaT_SPDC}
  \Delta T^2 =  \frac{2}{\sigma_p \varepsilon } \frac{\sum_{n=0}^\infty n \sinh^2 (\xi r_n) }{\sum_{m=0}^\infty \sinh^2 (\xi r_m) } \, .
\end{equation}

\begin{figure*}
  \includegraphics[width=.325\textwidth]{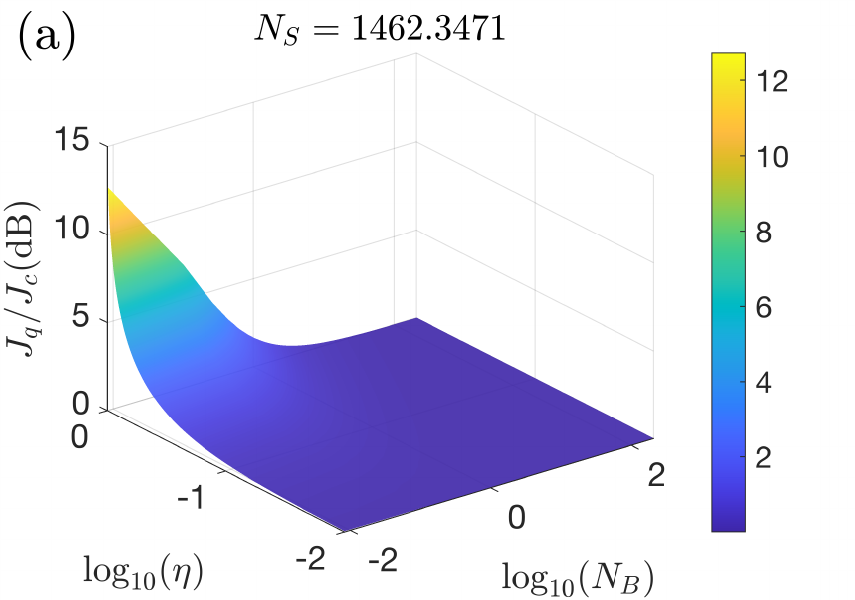}
  \includegraphics[width=.325\textwidth]{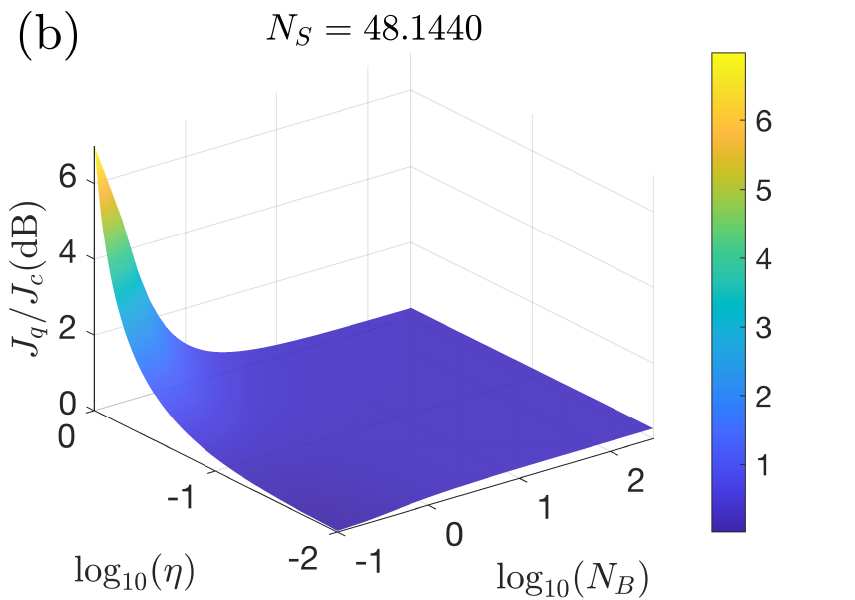}
  \includegraphics[width=.325\textwidth]{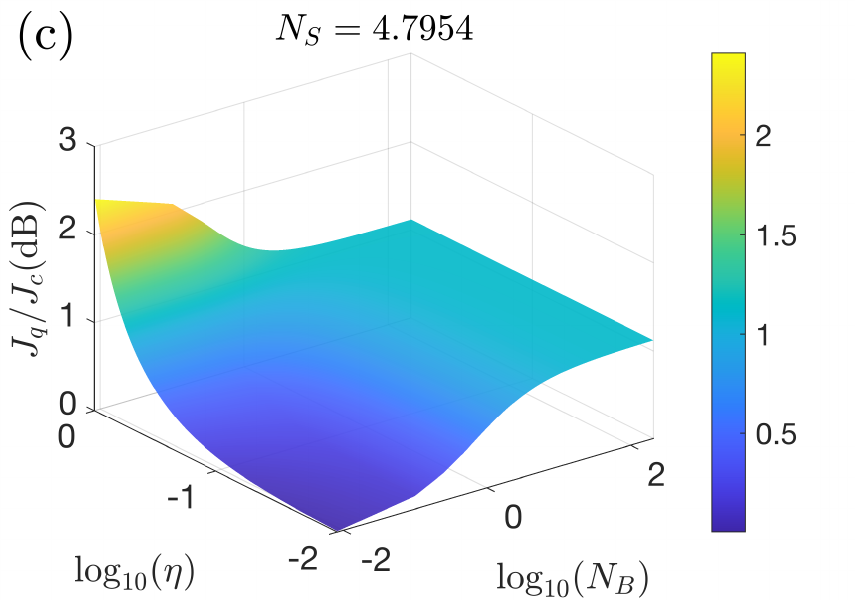}
  \includegraphics[width=.325\textwidth]{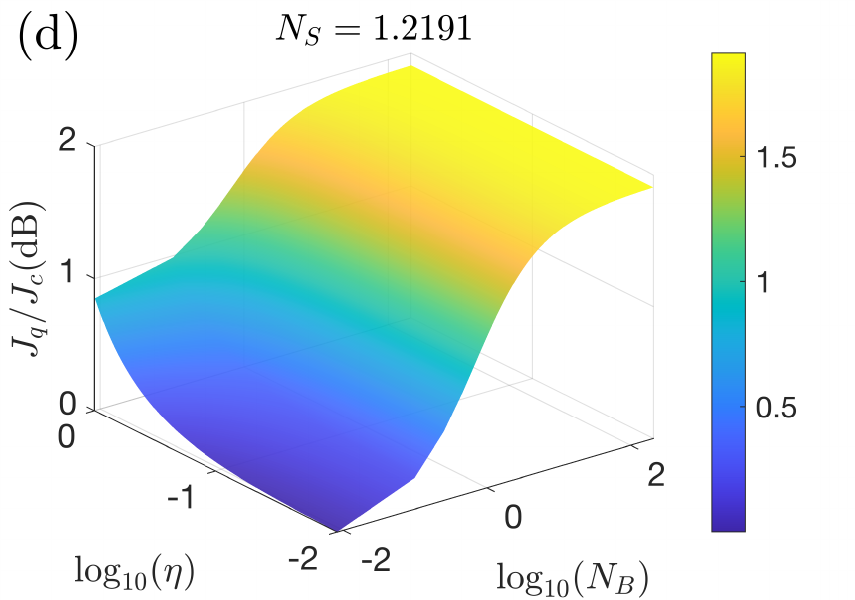}
  \includegraphics[width=.325\textwidth]{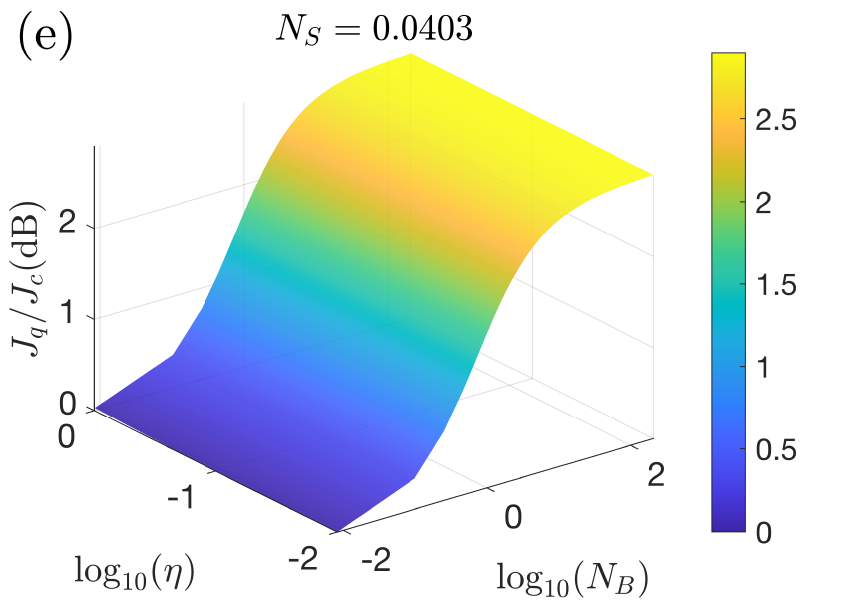}
  \includegraphics[width=.325\textwidth]{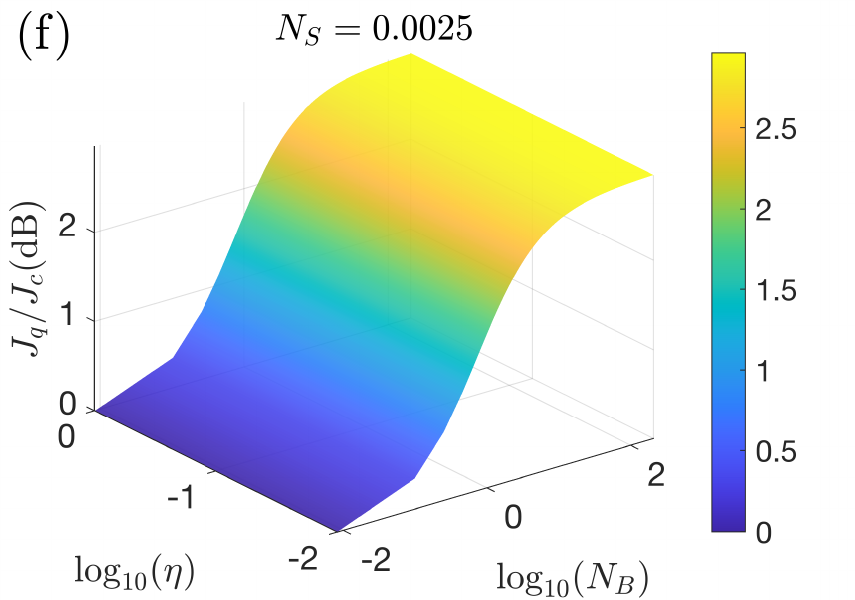}
  \caption{Quantum advantage, expressed by the ratio $J_q/J_c$, between the QFI for the quantum and classical strategies, as a function of $\eta$ and $N_B$ for different SPDC configurations.}
  \label{fig:JqJcratio_vs_eta_NB}
\end{figure*}

%%%%%%%%%%%%%%%%%%%%%%%%%%%%%%%%%
\section{Results}
\label{sec:results}

\subsection{Figures of merit and comparison scheme}
We want to compare the precision in the estimation of the Doppler parameter $\mu$ obtainable from the two classes of states defined in previous section.
We will refer to the protocol employing a coherent state for the signal beam as classical Doppler radar (CDR), while to the protocol based on SPDC states as quantum Doppler radar (QDR).
We make the standard assumption that the idler beam of the QDR configuration is perfectly stored without incurring in any additional noise.
While standard, this ideal assumption may be challenging to implement, especially for microwaves~\cite{Torrome2024}.

This comparison is carried out using the QFI as the figure of merit.
More details on this quantity and on quantum estimation theory are relegated to Sec.~\ref{subsec:qest}.
For understanding the results it suffices to know that the QFI is related to the optimal mean-square error of an estimator $\tilde{\mu}$ in the limit of many repetitions $m \gg 1$ as $ (\Delta^2 \tilde{\mu})_\mathrm{opt} \approx (m J[\rho_{\mu}] )^{-1}$; a larger QFI means a lower the estimation error.
The QFI only depends on the received quantum state of the radiation, the corresponding error amounts to an optimization over all measurements and classical estimators.
While optimal measurements may be hard to implement, and may also require adaptive steps in the estimation procedure~\cite{Barndorff-Nielsen2000}, this quantity sets a fundamental lower bound on the precision that it is physically possible to achieve.
Notice that for the CDR, i.e. a classical state of the field, the optimal measurement saturating the QFI may include detection schemes that could be considered nonclassical, such as photon counting.

For the CDR we obtain the following closed-form expression for the QFI
\begin{equation}
  \label{eq:Jc}
  J_c = \frac{4\eta N_S }{ \mu^2 \left({2 N_B + 1} \right)} 
    \int \! dy  {\left( \frac{1}{2}f(y) + y \frac{d f(y)}{d y} \right)}^2,
\end{equation}
which coincides with the expression obtained in Ref.~\cite{Reichert2022} for $N_B= 0$.
Details on this calculation are given in Sec.~\ref{subsec:QFI_coherent}.
In the limit $v/c \ll 1$ and $\Delta T \Delta \omega \, v/c \ll 1$ it can be approximated as
\begin{equation}
  \label{eq:Jcapprox}
  J_c \approx \frac{{ 4 {\omega_c}^2 \eta {N_S}  \Delta {T^2}}}{{\mu _0^2\left( {2{N_B} + 1} \right)}},
\end{equation}
we recall that for a coherent state $N_S = \alpha^2$, see Eq.~\eqref{eq:NS_coh}.
This regime is valid for the parameter regimes we consider and will be used in the following numerical calculations.

The QFI of the QDR will be denoted as $J_q$, but we have not found a simple closed-form expression.
Since the SPDC state is Gaussian and remains Gaussian after the Doppler encoding and noise, we take advantage of the compact description of Gaussian states in terms of first and second statistical moments, reviewed in Sec.~\ref{subsec:Gaussian}.
For Gaussian states, the QFI can be evaluated in terms of the first and second statistical moments~\cite{Monras2013,Safranek2019,Serafini2023}.
Concretely, numerical calculations are performed by truncating the number of Schmidt modes describing the SPDC, details on this calculation are in Sec.~\ref{subsec:QFI_TMSV}.

To ensure a fair comparison, we fix the time duration $\Delta T$ and the average photon number in the signal beam $N_S$ to have the same value.
The expressions for the CDR, i.e. a coherent state, are in Eq.~\eqref{eq:NS_coh} and Eq.~\eqref{eq:DeltaT_coh}, while for the QDR, i.e. a SPDC state, they are in Eq.~\eqref{eq:NS_SPDC} and Eq.~\eqref{eq:DeltaT_SPDC}. 
We also impose that the transmitted signals of both the CDR and QDR have the same central frequency $\omega_c$, which for the QDR based on SPDC means choosing a pump frequency $\omega_p = 2 \omega_c$.
This also ensures that they experience the same number of thermal photons $N_B$ as a function of the temperature.

The explicit physical parameter values used for the numerical comparison are: target speed $v=100 \mathrm{m}/\mathrm{s}$, and signal central frequency $\omega_c = 2 \pi c/\lambda$ with a wavelength in the microwave domain $\lambda = 6 \pi \mathrm{cm} \approx 18.8 \mathrm{cm}$.
Unless otherwise specified, we fix the pump bandwidth for the SPDC light to $\sigma_p =\omega_p/100$, where the pump central frequency is $\omega_p = 2 \omega_c $, and we work in the regime where the number of effective Schmidt modes is low, setting the phase matching bandwidth to 3 times the pump bandwidth, $\varepsilon = 3 \sigma_p$, corresponding to $K = 1.667$.
In this case, 5 Schmidt modes suffice to calculate QFI numerically.

\subsection{QFI comparison}

To concretely analyze the two strategies, we show the ratio $J_q / J_c$ between the QFI of the QDR and that of the CDR.
In absence of loss and thermal noise this ratio can scale linearly with the mean number of signal photons, since Heisenberg scaling for Doppler estimation can be reached~\cite{Reichert2022,Reichert2024}.
Even if this dramatinc difference in scaling is lost in presence of noise, a useful advantage can still be found.
In this section we present numerical results, highlighting the extent and scope of this advantage.

According to the results shown in Fig.~\ref{fig:JqJcratio_vs_eta_NB}, the following conclusions can be drawn:
\begin{enumerate}
 \item In panels (a) and (b) we show results for $\xi > K$, so that the average photon number of QDR is high $N_S \gg 1$.
 Quantum advantage is only present under low loss ($\eta \approx 1$) and low thermal noise ($N_B \to 0$) and in this low-noise regime, higher squeezing increases the quantum advantage.
  % (the higher the squeezing factor, the higher the signal-to-noise ratio), and the quantum advantage is also significant.
 However, under high loss and high thermal noise conditions, the QDR in this configuration exhibits essentially no advantage compared to the classical strategy.
\item In panel (c) we show results for $\xi = K$, corresponding to $N_S = 4.8$.
%  in panel (c).
% , corresponding to the ``mixed regime'' mentioned above.
Under these conditions, when the channel loss is low, the quantum advantage of QDR decreases with increasing thermal noise; in contrast, when the channel loss is high, the quantum advantage increases with increasing thermal noise.
It can be seen that the QDR in this configuration exhibits different characteristics under different signal conditions.
\item In panels (d), (e) and (f) we show results for $\xi < K$, corresponding to small values of $N_S$.
In this regime of low signal intensity, the quantum advantage of QDR increases with increasing thermal noise loss.
Especially when $\xi \ll K$, meaning $N_S \ll 1$, it can be observed from panels (e) and (f) that for strong thermal noise the quantum advantage can still reach 3 dB, indicating that the QDR behaves similarly to a QI protocol.
\end{enumerate}

\begin{figure*}
  \includegraphics[width=.325\textwidth]{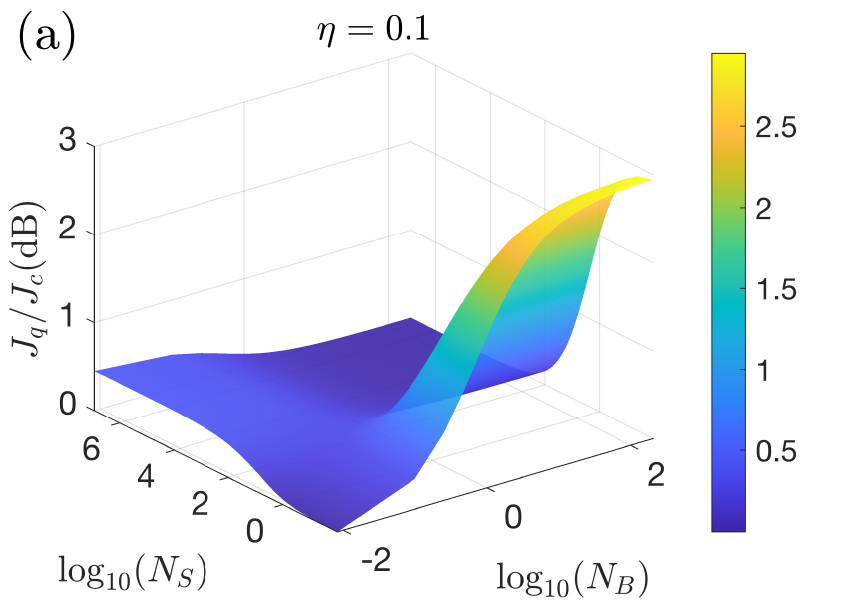}
  \includegraphics[width=.325\textwidth]{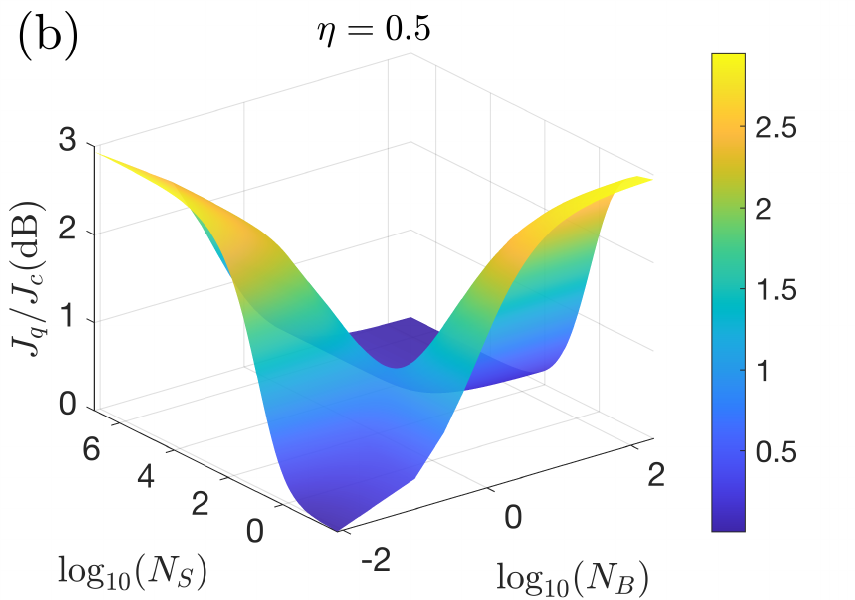}
  \includegraphics[width=.325\textwidth]{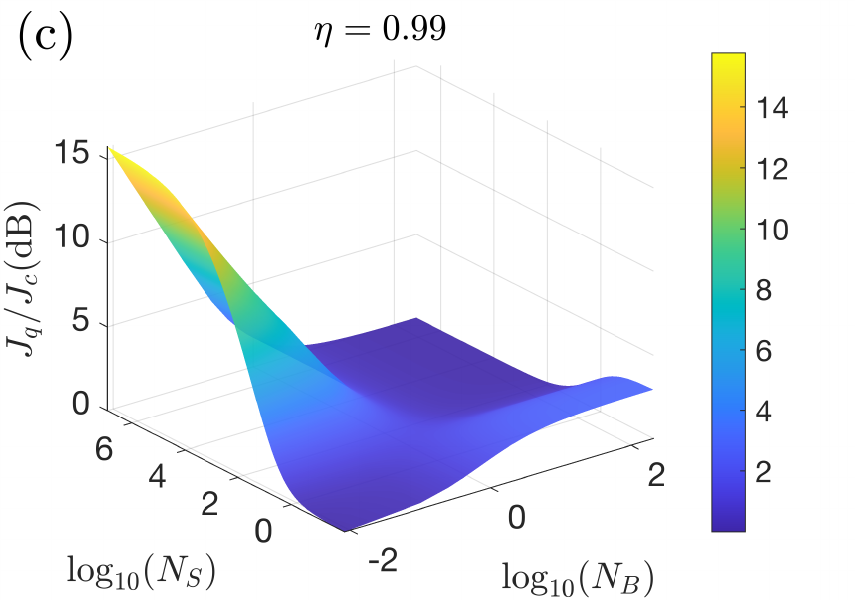}
  \caption{Quantum advantage, expressed by the ratio $J_q/J_c$, between the QFI for the quantum and classical strategies, as a function of $N_S$ and $N_B$ for different values of the transmissivity $\eta$.}
  \label{fig:JqJcratio_vs_NS_NB}
\end{figure*}

% \subsubsection{As a function of $N_S$ and $N_B$}

Having established that the conditions for quantum advantage of the entangled QDR are similar to those of QI, in Fig.~\ref{fig:JqJcratio_vs_NS_NB} we further analyse the dependence on the parameters $N_S$ and $N_B$, for a few exemplary values of $\eta$.
% This section analyzes the scope and conditions of quantum advantage that entangled quantum Doppler radar can achieve through the two main parameters that affect the quantum advantage of quantum illumination radar, 
% namely the average photon number $N_S$ of the radar signal source and the average photon number $N_B$ of thermal noise, so as to have a clearer understanding of its characteristics.
% As shown in Figure 3, it shows the relationship between the quantum advantage of QDR and the average number of photons in its signal source $N_S$ and the average number of photons in thermal noise $N_B$ under different link losses.
We can see see that regardless of the loss $\eta$, the QDR always shows a quantum advantage of up to about 3 dB in the regime $N_S \ll 1$ and $N_B \gg 1$, confirming the previous observation. 
In the opposite regime $N_S \gg 1$ and $N_B \ll 1$, the quantum advantage is instead related to the specific value of $\eta$: it is high for low losses and decreases as loss increases.
% and becomes relevant in the low loss scenario.
% % When the link loss is low, it can obtain a more significant quantum advantage.
% As the link loss increases, its quantum advantage gradually weakens.
% In the above evolution process, there are also some transition areas, which will not be analyzed in detail here.
% These analysis results provide a reference for us to further analyze the SPDC state signals in which band and parameter configuration are used in QDR, and how much advantages can be exerted when working in which scenarios.

%%%%%%%%%%%%%%%%%%%%%%%%%%%%%%%%%
\section{Conclusions and outlooks}
\label{sec:conclusions}

In this paper we have shown that a QI scheme has the potential to be advantageous also in the context of a Doppler radar for estimating the speed of a moving target.
While our results show such potential, similar to the original results in QI~\cite{Tan2008}, the main outstanding problem is to find a realistic detection schemes that can saturate the QFI.
Recently introduced detectors may prove useful for the task ~\cite{Angeletti2023,Reichert2023}, even though the intrinsic multimode aspect of the Doppler radar makes their analysis less straight-forward than in the standard QI scenario.

The analysis provided in this paper relies on local quantum estimation theory, which is generally valid only in the limit of many identical experiment repetitions.
Since this regime is not always appropriate for practical applications, it would be interesting to tackle the problem in terms of global quantum estimation theory.
Ref.~\cite{Zhuang2022} used Ziv-Zakai bounds to analyze range estimation, but other approaches may be possible.
An intriguing possibility would be to apply framework of Ref.~\cite{Rubio2022} for quantum scale estimation, since the Doppler parameter acts as a rescaling of frequencies.

Remaining in the QFI approach, an interesting and challenging question is to find the optimal probe state for a Doppler radar, similarly to what has been done for QI in the target detection task~\cite{DePalma2018b,Bradshaw2021}.
Alternatively, it would be useful to find channel bounds on the QFI which do not depend on particular choices for the probe state.
This approach has proven to be of great value for noisy quantum metrology with finite-dimensional systems~\cite{Demkowicz-Dobrzanski2012,Kurdzialek2023a}.
A similar complete theory for infinite-dimensional quantum channels is still missing, but several bounds for specific classes of Gaussian channels exist~\cite{Gagatsos2017,Pirandola2017,Nair2018,Nair2020,Nair2023}.

While the advantage of QI has been shown to often be impractical due real-world nonidealities, multiplexing techniques have recently been proposed to overcome some of these limitations~\cite{Zhao2024c}.
These ideas can likely be adapted to the quantum Doppler radar.
More in general, connections between target detection and parameter estimation have been shown in the context of estimating the reflectivity of the target (i.e. the loss parameter)~\cite{Sanz2017,Noh2022,Zhong2023}, but our results suggest there may be a more general connection.
The framework of Ref.~\cite{Meyer2023a} which puts parameter estimation and hypothesis testing on the same footing may be useful to explore this connection.

%%%%%%%%%%%%%%%%%%%%%%%%%%%%%%%%%
\section{Methods}
\label{sec:methods}

\subsection{Quantum estimation theory}
\label{subsec:qest}
For a family of states $\rho_{\theta}$ with a smooth parametric dependence on the real parameter $\theta$, the QFI is
\begin{equation}
  J[ \rho_{\theta} ] = \Tr \left[ L_\theta^2 \rho_{\theta} \right] \qquad  2 \frac{ d \rho_{\theta}}{d\theta} = \{ L_\theta, \rho_{\theta} \}
\end{equation}
where $L_\theta$ is the symmetric logarithmic derivative operator.
It is a generalization of the classical Fisher information for a probability distribution $F[p(x | \theta)]= \int dx p(x|\theta) \left( d \log p(x|\theta) / d\theta \right)^2 $.
In quantum mechanics, the probability distribution $p(x|\theta) = \Tr[ \rho_{\theta} \Pi_x ]$ is obtained from the state via the Born rule through a parameter-independent POVM (i.e. $\Pi_x \geq 0$ and $\int dx \Pi_x = \id$).
The QFI is attainable~\cite{Paris2009}, meaning that there exists at least a POVM with classical Fisher information equal to the QFI, i.e. $J[ \rho_{\theta} ] = \max_{\mathrm{POVMs}} F[p(x | \theta)]$.
For any unbiased estimator $\tilde{\theta}$ of the paramter $\theta$ the classical and quantum Cramér-Rao theorems set a lower bound on the variance~\cite{helstrom1976quantum,Holevo2011b,Paris2009}
\begin{equation}
  \Delta \tilde{\mu}^2 \geq \frac{1}{m F [ p ( x | \theta) ]} \geq \frac{1}{m J [ \rho_{\theta} ]}, 
\end{equation}
where $m$ is the number of identical repetitions of the experiment, or equivalently the number of identical copies of $\rho_{\theta}$ used in the experiment.
While it is generally impossible to find unbiased estimators for finite $m$, the bound has a deep meaning because it can be saturated asymptotically for large $m$, even though its saturation may require a two-step procedure when the optimal measurement depends on the true parameter value~\cite{Barndorff-Nielsen2000}.
Regardless of practical considerations, the QFI has a fundamental meaning in that it represents the maximal amount of information extractable about the parameter $\theta$ from a large number of identical copies of the state $\rho_{\theta}$.

\subsubsection{Dependence of the QFI on the parameter value}
\label{subsec:QFIparam_dep}

The fact that the QFI is a local quantity and it is invariant under parameter independent unitaries, leads to simplifications.
More specifically, this symmetry often allows one to compute the QFI for a single true value of the parameter and then obtain it for all the other values.
For example, for a phase parameter $\theta$ encoded as $\rho_\theta = e^{-i \theta \hat{G}} \rho_0 e^{i \theta \hat{G}}$ the QFI does not depend at all on the true value $\theta = \theta_0$.
This happens because one can think of applying the inverse unitary $e^{i \theta_0 \hat{G}}$ after parameter encoding (before measurement), i.e. the state $ \tilde{\rho}_{\theta}  = e^{i \theta_0 \hat{G}} \rho_\theta e^{-i \theta_0 \hat{G}}  = e^{-i(\theta - \theta_0 ) \hat{G}} \rho_0e^{-i(\theta - \theta_0 ) \hat{G}}$ has the same QFI as $\rho_\theta$.
Thus, since the QFI depends only on the state and its derivative evaluated at the true value, i.e. $\tilde{\rho}_{\theta = \theta_0} = \rho_0$ and $d \tilde{\rho}_{\theta} / d\theta |_{\theta=\theta_0} = -i [ \hat{G} , \rho_0]$, the result does not depend on the parameter value $\theta_0$.

For a phase parameter we have $U_{\theta_1}U_{\theta_2} = U_{\theta_1 + \theta_2}$, while for the Doppler parameter the composition rule is $U_{\mu_1} U_{\mu_2} = U_{\mu_1 \mu_2} $ (up to an additional $\mu$-independent phase shift, which can always be compensated and does not affect QFI calculations).
Therefore, the QFI depends on the true value of the parameter, but this dependence is simple and amounts to as a rescaling of the parameter. 
More explicitly, the QFI can be computed by considering the unitary $U_{1/\mu_0}U_{\mu}=U_{\mu/\mu_0}$, which amounts to a rescaling of the estimated parameter meaning that $J_q[ \rho_{\mu} ] |_{\mu = \mu_0} = J_q[ \rho_{\mu/\mu_0} ] |_{\mu=\mu_0} = \frac{1}{\mu_0^2} J_q[\rho_{\mu}] |_{\mu = 1} $.
Contrary to the rest of the paper, in the paragraph above we have explicitly made a distinction between the fixed true value of the parameter $\mu_0$, while $\mu$ is treated as a variable.
In the main text, we use a less rigorous (but standard) notation in which the two are conflated, in which the above argument becomes simply 
\begin{equation}
  J \left[ \rho_{r,\mu} \right] = \frac{1}{\mu^2} J \left[ \rho_{r,\mu=1} \right].
\end{equation}
Effectively, this means that it is enough to work around the value $\mu = 1$.
Interestingly, for the Doppler parameter the quantum signal-to-noise ratio~\cite{Paris2009} $\mu^2 J \left[ \rho_{r,\mu} \right] $ (which upper bounds the signal-to-noise ratio of an unbiased estimator $\mu^2 / \Delta \tilde{\mu}^2$) is independent of the true parameter value, thus respecting the underlying symmetry of the model.
The class of estimation problems with this symmetry is known as scale estimation~\cite{Rubio2022}.

\subsection{Gaussian states}
\label{subsec:Gaussian}

We only give a brief introduction, introducing the notation and concepts needed for the QFI calculation.
More details on Gaussian states and dynamics can be found in Refs.~\cite{Serafini2023,Weedbrook2012,Adesso2014}; we follow closely the notation of Ref.~\cite{Serafini2023}.

Given a finite collection of $M$ bosonic modes satisfying $[ \hat{A}_j, \hat{A}_k^\dag ]= \delta_{jk}$ and $[ \hat{A}_j, \hat{A}_k]=0$, $[ \hat{A}_j^\dag, \hat{A}_k^\dag]=0$, the corresponding hermitian position and momentum operators are defined as $\hat{Q}_j = ( \hat{A}_j+\hat{A}^\dag_j)/\sqrt{2} $ and $\hat{P}_j = - i ( \hat{A}_j+\hat{A}^\dag_j)/\sqrt{2} $ and satisfy the commutation relations $[\hat{Q}_j, \hat{P}_k] = i \delta_{jk}$.
We introduce the column vector of position and momentum operators $\vec{\hat{R}} = [ \hat{Q}_1 , \hat{P}_{1}, \cdots, \hat{Q}_M , \hat{P}_M ]^T $, so that the commutation relations in vectorial form can be written as $[ \vec{\hat{R}} , \vec{\hat{R}}^T ] = \Omega$, where the commutator of column and row vectors of operators should be taken as an outer product $[ \vec{\hat{R}} , \vec{\hat{R}}^T ]_{jk} = [ \hat{R}_j , \hat{R}_k ] $ (and analogously for anticommutators); the matrix $\Omega$ is the symplectic form 
\begin{equation}
  \label{eq:Sympl_form}
  \Omega  = \bigoplus \limits_{k = 1}^M \, \Omega_1 , \qquad 
  \Omega_1   = 
  \begin{pmatrix}
    0&1\\
    {- 1}&0
  \end{pmatrix}.
\end{equation}

A Gaussian state $\rho_G$ is a thermal or ground state of a Hamiltonian at most quadratic in the operators $\hat{R}_j$, or equivalently a state with a Gaussian Wigner function~\cite{Serafini2023}.
This class of states can be compactly defined in terms of a $M$-dimensional first moement vector $\vec{\bar{R}}$ and $2M{\times}2M$ covariance matrix $\vec{\sigma}$
\begin{equation}
  \vec{\bar{R}} = \Tr \left[ \rho_G \vec{\hat{R}} \right], \; \vec{\sigma} = \Tr \left[ \rho_{G} \left\{ (\hat{\vec{R}} - \vec{\bar{R}}), (\hat{\vec{R}} - \vec{\bar{R}})^T  \right\} \right].
\end{equation}
In this convention, the vacuum state and all coherent state have a covariance matrix equal to the identity: $\vec{\sigma}_c = \id_{2M}$.
A multimode coherent state generated by the displacement operator $e^{ \sum_{j=1}^M \alpha_j \hat{A}_j^\dag - \alpha_j^* \hat{A}_j}$ has first moments $\vec{\bar{R}}= \sqrt{2} [ \Re(\alpha_1) , \Im(\alpha_1), \dots, \Re(\alpha_M) , \Im(\alpha_M)  ]^T$
A multimode thermal state with the same number of thermal excitations $N_B$ for all modes, has a covariance matrix proportional to the identity, but larger: $\vec{\sigma}_{\mathrm{th}} = (2 N_B + 1) \id_{2M}$, and zero first moments.
Finally, a two-mode squeezed vacuum (or twin-beam) state has zero first moments and a covariance matrix that is locally thermal on the two subsystems, but it has correlations in positions and anticorrelation in momentum
\begin{equation}
  \vec{\sigma}_{\mathrm{tb}} = 
  \begin{pmatrix} 
    S & 0 &  C & 0 \\
    0 & S & 0 & - C \\ 
    C & 0 & S & 0 \\ 
    0 & - C & 0 & S
  \end{pmatrix},
\end{equation}
with $S=2 N_S + 1$ and $C= 2 \sqrt{ N_S (N_S +1) }$.
The multimode state obtained from realistic SPDC in Eq.~\eqref{eq:H_SPDC_disc} is just an uncorrelated product of twin-beam states, thus it has a covariance matrix in direct sum form
\begin{equation}
  \label{eq:sigma_spdc}
  \vec{\sigma}_{\mathrm{spdc}}  = \bigoplus_{i=1}^{M} \begin{pmatrix} 
    S_i & 0 &  C_i & 0 \\
    0 & S_i & 0 & - C_i \\ 
    C_i & 0 & S_i & 0 \\ 
    0 & - C_i & 0 & S_i
  \end{pmatrix},
\end{equation}
where now $S_i=2 N_{S,i} + 1$, $C_i= 2 \sqrt{ N_{S,i} (N_{S,i} +1) }$ and $ N_{S,i} = \sinh^2( \xi r_i )$. as in Eq.~\eqref{eq:NS_SPDC}.

\subsubsection{Gaussian channels}

A Gaussian channel\footnote{Note that a quantum channel is a Schrödinger picture description of the evolution of a quantum state given a \emph{fixed} set of modes.
However, practically speaking, this description does not match physical scenarios in which the field is propagating and the modes at the sender and at the receiver's ends are different.
The two pictures can be reconciled; while this is probably known to experts, in Appendix~\ref{app:schrodinger_picture_propagating_fields} we provide a simple pedagogical argument.
}
is defined in terms of two matrices $X$ and $Y$, that act as follows on the first moments and covariance matrix~\cite{Caruso2008,Serafini2023}
\begin{align}
  \vec{\bar{R}} & \mapsto X \vec{\bar{R}}  \\ 
  \vec{\sigma} & \mapsto X \vec{\sigma} X^T  + Y,
\end{align}
and they have to satisfy $Y + i \Omega \geq i X \Omega X^T$.
When the channel represents a unitary transformation then $Y=0$ and $X=S$, where $S$ is a symplectic transformation satisfying $S\Omega S^T = \Omega$.
In particular, an $M$-mode thermal loss channel (where loss and thermal excitations are identical on all modes) has $X=\sqrt{\eta} \id_{2M} $ and $Y=(1-\eta)(2 N_B + 1) \id_{2M}$.

\subsubsection{QFI of Gaussian states}

For a parametric family of Gaussian states $\rho_G(\theta)$ the QFI can be evaluated just in terms of the parametric derivatives (denoted with a dot) of the first and second moments~\cite{Monras2013,Serafini2023,Safranek2019}\footnote{Here, we neglect cases in which some of the normal modes are pure (i.e. one or more symplectic eigenvalues are $1$) and the matrix inversion is not well-defined.
The problem can be solved through regularization procedures~\cite{Safranek2019}, but for $N_B > 0$ this is never needed for our model.
}:
\begin{align}
    & J[\rho_G(\theta)] =  \,  2 \, \left(\partial_\theta \bar{\vec{R}} \right)^T  \vec{\sigma}^{- 1}\left(\partial_\theta \bar{\vec{R}} \right) \label{eq:gaussianQFIfirst} \\ 
    & + \frac{1}{2}\operatorname{vec}{\left( \partial_\theta {\vec{\sigma}} \right)^T}{\left( { \vec{\sigma}  \otimes \vec{\sigma}  - \Omega  \otimes \Omega } \right)^{ - 1}}\operatorname{vec}\left( \partial_\theta {\vec{\sigma}}\right). \label{eq:gaussianQFIcm}
\end{align}
Eq.~\eqref{eq:gaussianQFIfirst} is the contribution from the first moments, and corresponds to the formula of the classical Fisher information for a parameter in the mean values of a Gaussian distribution.
Eq.~\eqref{eq:gaussianQFIcm} is the contribution from the  covariance matrix; the operation $\operatorname{vec}(M)$ is the vectorization of a matrix, i.e. the vector obtained by stacking the columns of the matrix on top of each other.
If the symplectic diagonalization of $\vec{\sigma}$ is known, alternative expressions for the covariance matrix contribution~\eqref{eq:gaussianQFIcm} are also available~\cite{Monras2013,Serafini2023,Safranek2019}.

\subsection{QFI calculations}
\label{subsec:QFIcalc}

Both the CDR and the QDR can be described using Gaussian states, thus Eqs.~\eqref{eq:gaussianQFIfirst} and~\eqref{eq:gaussianQFIcm} can be applied.
However, we need to pay attention to choosing the appropriate basis of modes when performing the calculation.
A similar scenario is considered in Ref.~\cite{Sorelli2024}, with the assumption that the modes that are not populated by the parameter-dependent quantum state are in the vacuum.
Those results are not immediately applicable to our multimode model with identical and independent thermal noise acting on all the modes.
While the basis of discrete modes defined in Sec.~\ref{sec:model} is infinite-dimensional, it can be truncated to a finite value for concrete calculations.
Therefore, while the matrices we are considering here should be in principle infinite dimensional, we work in the assumption number of modes will be eventually truncated to a finite value $M$.

We start from the Doppler unitary transformation of discrete modes in Eq.~\eqref{eq:Doppler_Heisenberg_discrete}.
The unitary transformation in terms of $\hat{A}_j$ operators corresponds to a symplectic transformation at the level of $\hat{R}_j$ operators:

\begin{equation}
  \scriptsize
S_\mu =
\begin{pmatrix}
\operatorname{Re}(\mathcal{U}_{\mu,11}) & -\operatorname{Im}(\mathcal{U}_{\mu,11}) & \dots & \operatorname{Re}(\mathcal{U}_{\mu,1M}) & -\operatorname{Im}(\mathcal{U}_{\mu,1M}) \\
\operatorname{Im}(\mathcal{U}_{\mu,11}) & \operatorname{Re}(\mathcal{U}_{\mu,11}) & \dots & \operatorname{Im}(\mathcal{U}_{\mu,1M}) & \operatorname{Re}(\mathcal{U}_{\mu,1M}) \\
\vdots & \vdots & \ddots & \vdots & \vdots \\
\operatorname{Re}(\mathcal{U}_{\mu,M1}) & -\operatorname{Im}(\mathcal{U}_{\mu,M1}) & \dots & \operatorname{Re}(\mathcal{U}_{\mu,MM}) & -\operatorname{Im}(\mathcal{U}_{\mu,MM}) \\
\operatorname{Im}(\mathcal{U}_{\mu,M1}) & \operatorname{Re}(\mathcal{U}_{\mu,M1}) & \dots & \operatorname{Im}(\mathcal{U}_{\mu,MM}) & \operatorname{Re}(\mathcal{U}_{\mu,MM})
\end{pmatrix}
\end{equation}
Overall, the channel defined by Eq.~\eqref{eq:Doppler_mode_transf_discrete} corresponds to a Gaussian quantum channel with the following parameters
\begin{equation}
  \label{eq:XYGaussian_channel_DopplerThermal}
  X = \sqrt{\eta} S_\mu \quad Y= (2 N_B + 1 - \eta ) \id_{2M},
\end{equation}
we remind that the multiplicative factor in $Y$ comes from the rescaling of the thermal excitations by $(1-\eta)$, as explained in Sec.~\ref{sec:model}.
We see that the Doppler parameter affects the state only through the symplectic matrix $S_\mu$.

Following the argument in Sec.~\ref{subsec:QFIparam_dep}, for the QFI calcualtion we can work around the value $\mu=1$ without loss of generality.
The main quantity left to evaluate is thus the parametric derivative of the unitary transformation between the discrete modes
\begin{align}
  \label{eq:dUdiscrete}
   \frac{d}{d \mu} & \mathcal{U}_{\mu,k j} \mid_{\mu = 1} =  \int d \omega \psi_k(\omega) \frac{d}{d \mu} \left( {-\mu ^{1/2}} \psi_j( \mu \omega ) \right) \mid_{\mu = 1}  \\
 =& - \int d \omega \psi_k(\omega)  \left( \frac{1}{2}\psi_j( \omega ) + \omega \frac{d}{d\omega} \psi_j (\omega )\right).
 \end{align}

\subsubsection{Coherent state}
\label{subsec:QFI_coherent}

For a coherent state, we use a mode basis in which the first element is $\Psi_1(\omega) = f(\omega)$, i.e. the function defining the trasmitted coherent state and the second mode of the basis is proportional to the parametric derivative $\frac{d}{d \mu} \left( \mu^{-1/2} f(\omega/\mu) \right) \mid_{\mu=1}$, which taking into account the normalization becomes
\begin{equation}
\Psi_2(\omega) = \frac{1}{\sqrt{\mathcal{N}}} \left( \frac{1}{2} f(\omega) + \omega \frac{d f(\omega)}{d \omega} \right),
\end{equation}
with $\mathcal{N} = \int d\omega \left(\frac{1}{2} f(\omega) + \omega \frac{d f(\omega)}{d \omega} \right)^2$.
Notice that the orthogonality follows from the fact that $f(\omega) \in \mathbb{R}$ is square integrable and must vanish at $\pm \infty$; it also follows from taking the derivative of the identity $ \int \frac{d \omega}{\mu} f(\omega/\mu)^2 = 1$.
We will not need the rest of the basis elements for the calculation, but it is enough to know that they will orthogonal to both $\Psi_1$ and $\Psi_2$.
This property implies that in Eq.~\eqref{eq:dUdiscrete} only the matrix elements $\dot{\mathcal{U}}_{\mu,10}$ and $\dot{\mathcal{U}}_{\mu,10}$ are non-zero, so we obtain the first moments of the received state
\begin{equation}
  \partial_\mu \bar{\vec{R}}_r = \begin{pmatrix}0 \\ 0 \\ \sqrt{2 \mathcal{N}} \alpha \\ 0 \end{pmatrix}.
\end{equation}
Since the initial covariance matrix of a coherent state is the identity, applying the Gaussian channel described in Eq.~\eqref{eq:XYGaussian_channel_DopplerThermal} gives the covariance matrix of the received state $\vec{\sigma}_r = (2 N_B + 1 )\id_{2 M}$.
Finally, applying the first-moment contribution to the Gaussian QFI in Eq.~\eqref{eq:gaussianQFIfirst} we get to the expression reported in Eq.~\eqref{subsec:QFI_coherent}.

\subsubsection{SPDC state}
\label{subsec:QFI_TMSV}

Applying the thermal noise channel to the SPDC state gives a covariance matrix of the following form for the received state
\begin{equation}
  \label{eq:sigma_spdc_received}
  \vec{\sigma}_{\mathrm{spdc},r}  = \bigoplus_{i=1}^{M} \begin{pmatrix} 
     A_i & 0 &  \sqrt{\eta} C_i & 0 \\
    0 &  \eta A_i & 0 & - \sqrt{\eta} C_i \\ 
    \sqrt{\eta} C_i & 0 & S_i & 0 \\ 
    0 & - \sqrt{\eta} C_i & 0 & S_i
  \end{pmatrix},
\end{equation}
with $A_i = 2 \eta N_{S,i} + 2 N_B + 1 $.
For each pair of signal-idler Schmidt modes, the symplectic eigenvalues and transformation are the same as in Ref.~\cite{Tan2008}.

For the SPDC state there is a natural mode basis, defined by Schmidt decomposition in Eq.~\eqref{eq:Schmidt_dec}.
For the Gaussian JSA in Eq.~\eqref{eq:JSA_Gaussian} we have that the two families of polynomials for signal and idler are identical: 
\begin{equation}
  \label{eq:Schmidt_dec_Gauss}
f({\omega _S},{\omega _I}) = \sum\limits_{m = 0}^\infty  {{r_m}{\psi _m}({\omega _S} - {\omega _c}/2){\psi _m}({\omega _I} - {\omega _c}/2)}, 
\end{equation}
and they are Hermite-Gauss polynomials:
\begin{align}
  \label{eq:}
  {\psi _m}(\omega  - {\omega _c}/2) =& \sqrt s {\varphi _m}\left[ {s(\omega  - {\omega _c}/2)} \right] \\
  {\varphi _n}(\omega ) =& {({2^n}n!\sqrt \pi  )^{ - 1/2}}{H_n}(\omega ){e^{ - {\omega ^2}/2}}
\end{align}
where $s = \sqrt {\frac{2}{{\sigma_p \varepsilon }}} $ and $H_n(x) = (-1)^n e^{x^2}\frac{d^n}{dx^n}e^{-x^2}$ are the physicist's Hermite polynomials.
The functions $\varphi_n(y)$ are the harmonic oscillator eigenfunctions and satisfy the following properties
\begin{equation}
  \begin{aligned}
  \frac{d \varphi_n(y) }{d y} & = -\frac{ n+1}{2}  \varphi_{n+1}(y) + \frac{ n}{2}  \varphi_{n-1}(y) , \\
  y \varphi_n(y) & = \frac{ n+1}{2}  \varphi_{n+1}(y) + \frac{ n}{2}  \varphi_{n-1}(y) .
\end{aligned}
\end{equation}
These relations allow us to express the element $\omega \frac{d}{d\omega} \psi'_j (\omega )$ in Eq.~\eqref{eq:dUdiscrete} as a linear combination of functions $\psi_{k}$ with $k=j-2,j-1,j,j+1,j+2$.
The complete expression if lengthy and is not reported here, however it is straightforward to write all the relevant matrix elements.

From these expressions for $\mathcal{U}_{\mu}$ we can thus obtain the derivative of the symplectic matrix $\partial_\mu S_\mu$ and thus the derivative of the received density matrix 
\begin{equation}
\partial_\mu \vec{\sigma}_r = (\partial_\mu \tilde{S}_\mu)  \vec{\sigma}_{\mathrm{spdc},r} +  \vec{\sigma}_{\mathrm{spdc},r}  (\partial_\mu \tilde{S}_\mu)^T.
\end{equation}
Here we use the tilde to denote that the symplectic transformation acts only on the signal modes. 
Ordering the modes such that the signal modes come first and then the idler modes, this corresponds to a direct sum $\tilde{S}_\mu = S_\mu \oplus \id_{2M}$.
This is not the order used in Eqs.~\eqref{eq:sigma_spdc} and \eqref{eq:sigma_spdc_received}; it is immediate to switch from one order to the other by appropriate permutation matrices.
Finally, the QFI can be computed numerically from Eq.~\eqref{eq:gaussianQFIcm}, since the first moment vector is zero in this case.

\appendix 

%%%%%%%%%%%%%%%%%%%%%%%%%%%%%%%%%
\section{Comment on Heisenberg and Schrödinger picture for propagating quantum fields}
\label{app:schrodinger_picture_propagating_fields}

We point out a conceptual subtlety in translating from Heisenberg to Schrödinger picture in the context of quantum optics.
These ideas are likely already well-known to experts, but we find it instructive to clearly explain the situation.

We consider a linear input-output relation between bosonic creation and annihilation operators (formally, a Bogoliubov transformation):
\begin{equation}
  \begin{split}
    \vec{\hat{\alpha}}_\mathrm{out} \equiv \begin{bmatrix} \vec{\hat{a}}_{\mathrm{out}} \\ \vec{\hat{a}}^\dag_{\mathrm{out}} \\ \vec{\hat{b}}_{\mathrm{out}} \\ \vec{\hat{b}}^\dag_{\mathrm{out}}
    \end{bmatrix} =& 
    \hat{U}^\dag(\vec{\hat{\alpha}}_\mathrm{in}) 
    \vec{\hat{\alpha}}_\mathrm{in} 
    \hat{U}(\vec{\hat{\alpha}}_\mathrm{in}) \\ 
    &  = \vec{M}_U \vec{\hat{\alpha}}_\mathrm{in} 
    % \quad \vec{\hat{\alpha}}_\mathrm{in} 
    \equiv
    \vec{M}_U  \begin{bmatrix} \vec{\hat{a}}_{\mathrm{in}} \\ \vec{\hat{a}}^\dag_{\mathrm{in}} \\ \vec{\hat{b}}_{\mathrm{in}} \\ \vec{\hat{b}}^\dag_{\mathrm{in}}
    \end{bmatrix}
  \end{split}
\end{equation}
Note that we use the round brackets in $\hat{U}(\vec{\hat{\alpha}}_\mathrm{in})$ to denote the \emph{functional} dependence of the operator $\hat{U}$ on the bosonic input operators, so that, e.g., $\hat{U}(\vec{\hat{\alpha}}_\mathrm{out})$ represents a different operator, but with the same functional dependence on the bosonic operators in the round brackets.
Similarly, for the initial state $\rho_0(\vec{\hat{\alpha}}_\mathrm{in})$ denotes the functional form of the state in terms of the input Bosonic operators.
We will always assume that the functions are well-behaved so that $ U^\dag f(X) U = f(U^\dag X U)$ for a unitary operator $U$ (this should always hold for analytical functions expressible as power series, such as the exponential function that defines unitaries).

In Schrödinger picture we consider the partition $\mathcal{H}_{\vec{\hat{a}}_{\mathrm{in}}} \bigotimes \mathcal{H}_{\vec{\hat{b}}_{\mathrm{in}}}$ of the total Hilbert space\footnote{The notation $\mathcal{H}_{\vec{\hat{a}}_{\mathrm{in}}}$ should be self-explanatory, but formally it means that this is the Fock space built by the corresponding creation operators.}, i.e. a split between the Hilbert space of the initial modes $\vec{\hat{a}}_{\mathrm{in}}$ and the additional modes $\vec{\hat{b}}_{\mathrm{in}}$, which  will be considered to be inaccessible, i.e. an environment.
Tracing out the environment modes means tracing over the Hilbert space $\mathcal{H}_{\vec{\hat{b}}_{\mathrm{in}}}$, so that the accessible state is an operator acting only the Hilbert space $\mathcal{H}_{\vec{\hat{a}}_{\mathrm{in}}}$ defined by the input modes.
If the initial state $\rho_0(\vec{\hat{\alpha}}_\mathrm{in})$ is factorized across the bipartition $\mathcal{H}_{\vec{\hat{a}}_{\mathrm{in}}} \bigotimes \mathcal{H}_{\vec{\hat{b}}_{\mathrm{in}}}$, this procedure corresponds to applying a Gaussian quantum channel on the initial state, an operator acting on $\mathcal{H}_{\vec{\hat{a}}_{\mathrm{in}}}$ and this description is essentially a Stinespring dilation of the channel.

Now the crucial observation is that, from a physical point of view, in quantum optics the local operators available at the receiver are not $\vec{\hat{a}}_{\mathrm{in}}$, but $\vec{\hat{a}}_{\mathrm{out}}$, thus the relevant Hilbert space partition at the receiver's end is $\mathcal{H}_{\vec{\hat{a}}_{\mathrm{out}}} \bigotimes \mathcal{H}_{\vec{\hat{b}}_{\mathrm{out}}}$.
We want to find the form of the density operator $ \rho_{\mathrm{out}}(\vec{\hat{a}}_{\mathrm{out}})$ at the receiver: this operator acts on $\mathcal{H}_{\vec{\hat{b}}_{\mathrm{out}}}$ and it can be used to compute the expectation values of all local operators at the receiver, which we write as $f( \vec{\hat{a}}_{\mathrm{out}})$ for an arbitrary function $f$.
We can thus write
\begin{align*}
    % & \Tr_{\vec{\hat{a}}_{\mathrm{out}}} \left[ f( \vec{\hat{a}}_{\mathrm{out}}) \rho_{\mathrm{out}}(\vec{\hat{a}}_{\mathrm{out}}) \right] = 
    & \Tr_{\vec{\hat{a}}_{\mathrm{out}}} \left[ f( \vec{\hat{a}}_{\mathrm{out}}) \Tr_{{\vec{\hat{b}}_{\mathrm{out}}}} \left[ \rho_0 (\vec{M}_U^{-1}\vec{\hat{\alpha}}_{\mathrm{out}}) \right] \right] \\ 
    &= \Tr \left[ f( \vec{\hat{a}}_{\mathrm{out}}) \rho_0 (\vec{\hat{\alpha}}_{\mathrm{in}}) \right]  \\ 
    &=  \Tr \left[ f( \vec{\hat{a}}_{\mathrm{in}}) \hat{U}(\vec{\hat{\alpha}}_\mathrm{in}) \rho_0 ( \vec{\hat{\alpha}}_{\mathrm{in}}) \hat{U}^\dag(\vec{\hat{\alpha}}_\mathrm{in}) \right] \\
    & = \Tr_{\vec{\hat{a}}_{\mathrm{in}}} \left[ f( \vec{\hat{a}}_{\mathrm{in}}) \Tr_{{\vec{\hat{b}}_{\mathrm{in}}}} \left[ \hat{U}(\vec{\hat{\alpha}}_\mathrm{in}) \rho_0 ( \vec{\hat{\alpha}}_{\mathrm{in}}) \hat{U}^\dag(\vec{\hat{\alpha}}_\mathrm{in})  \right]  \right], 
\end{align*}
where from the first to the second line we moved from Heisenberg to Schrödinger picture.
Since the identity between the initial and final expressions must hold for any $f$, and since the output Hilbert spaces are isomorphic to the input Hilbert spaces, we deduce that 
\begin{align}
    \rho_{\mathrm{out}}(\vec{\hat{a}}_{\mathrm{out}}) &\equiv \Tr_{{\vec{\hat{b}}_{\mathrm{out}}}} \left[ \rho_0 (\vec{M}_U^{-1}\vec{\hat{\alpha}}_{\mathrm{out}}) \right] \\ 
    & = \Tr_{{\vec{\hat{b}}_{\mathrm{out}}}} \left[ \hat{U}(\vec{\hat{\alpha}}_\mathrm{out}) \rho_0 ( \vec{\hat{\alpha}}_{\mathrm{out}}) \hat{U}^\dag(\vec{\hat{\alpha}}_\mathrm{out})  \right].
\end{align}
This means that the state at the receiver is formally identical to the Schrödinger picture evolved state, up to an isomorphism between the input and output Hilbert spaces.

\begin{figure}
  \includegraphics[width=.46\textwidth]{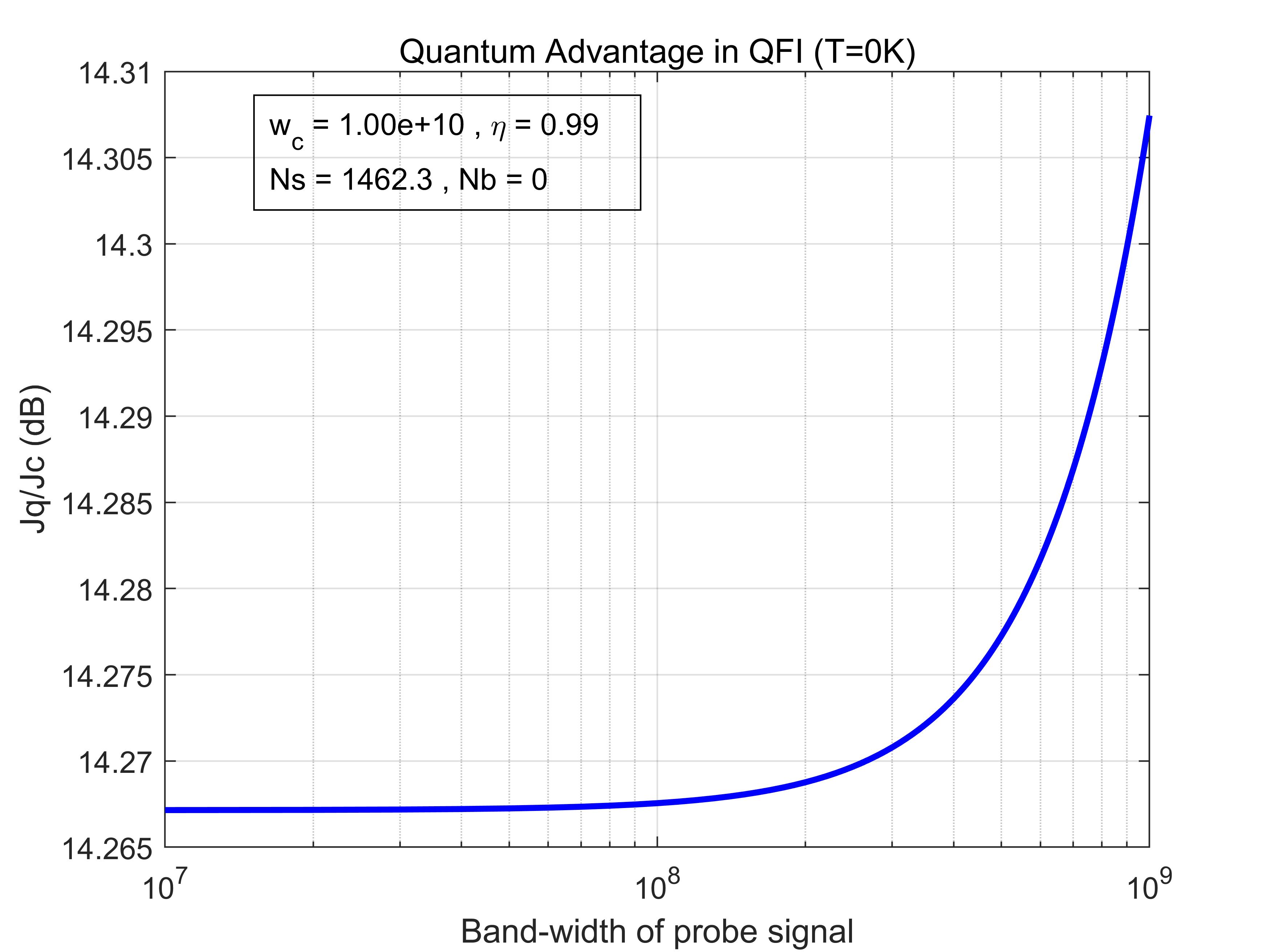}
  \includegraphics[width=.46\textwidth]{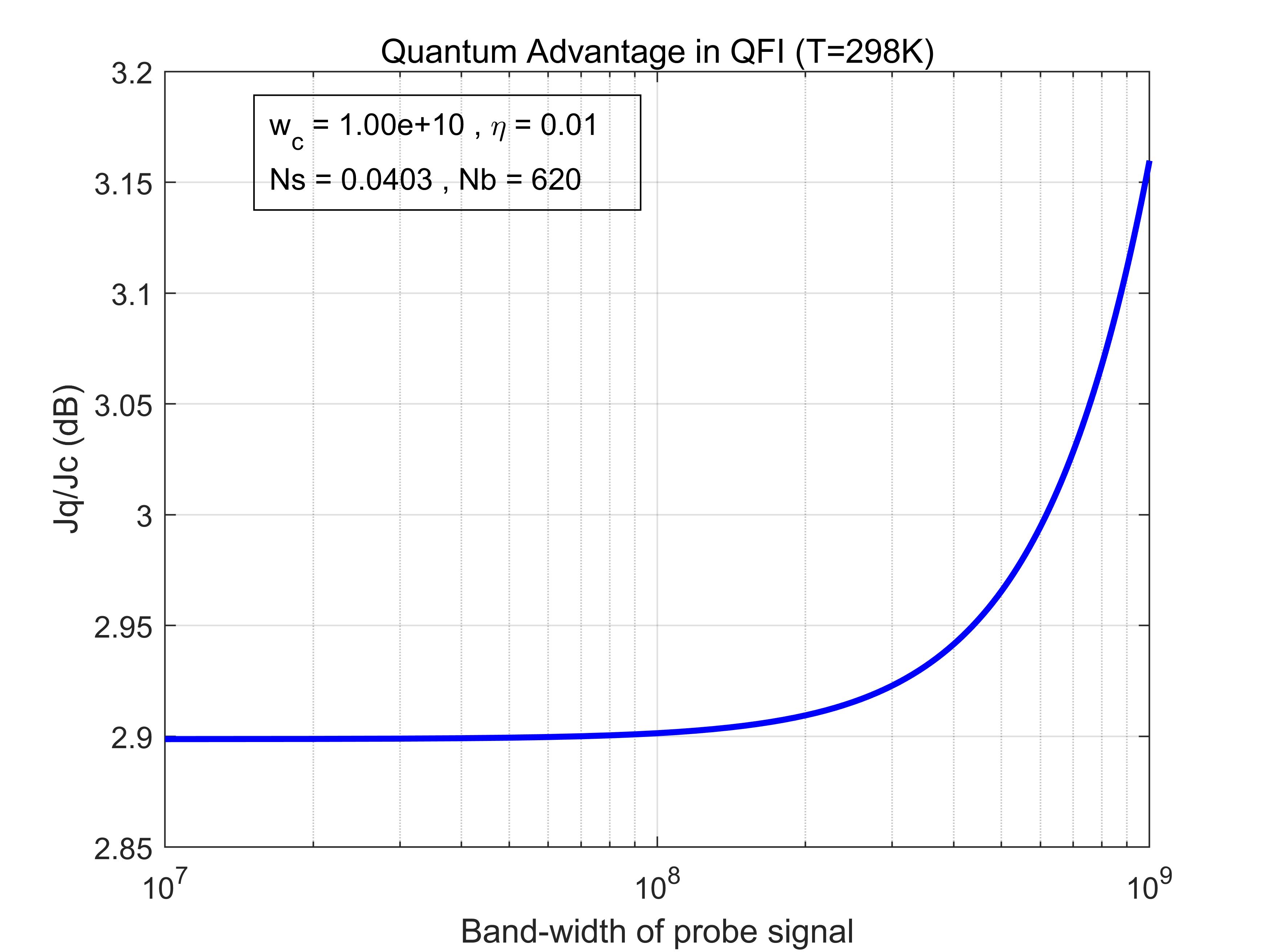}
  \caption{Ratio $J_q/J_c$ as a function of the pump bandwidth for different SPDC configurations}
  \label{fig:JqJcratio_vs_bandwidth}
\end{figure}
\section{Impact of pump bandwidth on the quantum advantage}

In the main text we have shown that the QDR can have obvious quantum advantages in the following two scenarios: (a) low noise (both loss and thermal noise) and a large average number of signal photons; (b) high thermal noise and low average number of signal photons.
From the approximated QFI of the CDR in Eq.~\eqref{eq:Jcapprox} it is clear that the time duration of the signal also affects the accuracy of the radar in measuring the Doppler shift induced by the target.

Generally speaking, a wider spectral bandwidth is related to a worse ability to resolve different frequencies, and thus a lower Doppler measurement accuracy, as quantified by the QFI.
Conversely, a narrow bandwidth should give a higher the measurement accuracy.
In Fig.~\ref{fig:JqJcratio_vs_bandwidth} we study the influence of pulse bandwidth $\sigma_p$, on the quantum advantage of the QDR.
The center frequency of the probe signal is set to $\omega_c = \omega_p/2 = 1010 \mathrm{Hz}$, which corresponds to a signal with a wavelength of $\lambda = 6 \pi \mathrm{cm}$ and the bandwidth $\sigma_p$ spans values from $\omega_c/1000$ to $\omega_c/10$, while $\varepsilon$ is fixed.
From these values we compute $N_S$ and $\Delta T$ and put the corresponding values to evaluate $J_c$ for the classical protocol.

Fig.~\ref{fig:JqJcratio_vs_bandwidth} shows that there is relatively weak increase of the quantum advantage as the pump bandwidth increases, for both scenarios (a) and (b) described above.
In particular, in panel (a) we see that for the scenario (a) the effect is very moderate and can essentially be neglected, confirming the observation made in Ref.~\cite{Reichert2022} for the noiseless $\eta=1$ scenario.
In panel (b), we see that for scenario (b) the enhancement can reach more than 0.25 dB, and thus it is not entirely negligible. 
However, we also remark that going to a large pump bandwidth may break the narrowband approximation under which the model was derived, a more refined model is needed for a proper analysis.

\bibliography{doppler_radar_biblio}

\end{document}